\documentclass[twocolumn,superscriptaddress,amsmath,amssymb,prl,aps,floatfix]{revtex4-1}

\usepackage{graphicx}
\usepackage{dcolumn}
\usepackage{bm}
\usepackage{multirow}
\usepackage{booktabs}
\usepackage{color}

\usepackage{longtable}
\usepackage{etoolbox}
\usepackage{url}
\usepackage{breakurl} 
\usepackage[breaklinks]{hyperref}
\apptocmd{\sloppy}{\hbadness 10000\relax}{}{}

\renewcommand{\v}[1]{\ensuremath{\mathbf{#1}}} 

\AtBeginDocument{
\heavyrulewidth=.08em
\lightrulewidth=.05em
\cmidrulewidth=.03em
\belowrulesep=.65ex
\belowbottomsep=0pt
\aboverulesep=.4ex
\abovetopsep=0pt
\cmidrulesep=\doublerulesep
\cmidrulekern=.5em
\defaultaddspace=.5em
}

\begin{document}


\title{Large Bulk Photovoltaic Effect and Spontaneous Polarization\\ of Single-Layer Monochalcogenides}

\author{Tonatiuh Rangel}
\thanks{These two authors contributed equally.}
\affiliation{Molecular Foundry, Lawrence Berkeley National Laboratory, Berkeley, California 94720, USA}
\affiliation{Department of Physics, University of California, Berkeley, California 94720, USA}

\author{Benjamin \surname{M. Fregoso}}
\thanks{These two authors contributed equally.}
\affiliation{Department of Physics, University of California, Berkeley, California 94720, USA}
\affiliation{Department of Physics, Kent State University, Kent, Ohio 44242, USA}

\author{Bernardo S. \surname{Mendoza}}
\affiliation{Centro de Investigaciones en \'Optica, Le\'on, Guanajuato 37150, M\'exico}

\author{Takahiro Morimoto}
\affiliation{Department of Physics, University of California, Berkeley, California 94720, USA}

\author{\\Joel E. \surname{Moore}}
\affiliation{Department of Physics, University of California, Berkeley, California 94720, USA}

\author{Jeffrey B. \surname{Neaton}}
\affiliation{Molecular Foundry, Lawrence Berkeley National Laboratory, Berkeley, California 94720, USA}
\affiliation{Department of Physics, University of California, Berkeley, California 94720, USA}
\affiliation{Kavli Energy Nanosciences Institute at Berkeley, Berkeley, California 94720, USA}


\begin{abstract}
We use a first-principles density functional theory approach to calculate the shift current and linear absorption 
of uniformly illuminated single-layer Ge and Sn monochalcogenides. We predict strong absorption in the visible 
spectrum and a large effective three-dimensional shift current ($\sim$100~$\mu$A/V$^2$), larger than has been 
previously observed in other polar systems. Moreover, we show that the integral of the shift-current tensor is 
correlated to the large spontaneous effective three-dimensional electric polarization ($\sim$1.9~C/m$^2$). Our 
calculations indicate that the shift current will be largest in the visible spectrum, suggesting that these 
monochalcogenides may be promising for polar optoelectronic devices. A Rice-Mele tight-binding model is used to 
rationalize the shift-current response for these systems, and its dependence on polarization, in general terms with 
implications for other polar materials
\end{abstract}

\maketitle


{\it Introduction}.-- The shift current is a dc current generated in a material under uniform 
illumination~\cite{Sturman1992,von_baltz_theory_1981,sipe_second-order_2000,Tan2016}, and gives rise to 
phenomena such as the bulk photovoltaic effect (BPVE)~\cite{Sturman1992}. A necessary condition for the BPVE in a 
material is the lack of inversion symmetry. Interestingly, in the BPVE, the resulting photovoltage is not limited 
by the band gap energy, and a junction or interface is not required to generate a current. These properties of the 
BPVE motivate great interest in the possible optoelectronic applications of  noncentrosymmetric systems, and they have 
been suggested to play a role in emerging functional materials, including hybrid halide 
perovskites~\cite{young_first-principles_2012,*young_first_2012,*zheng_first-principles_2015,*young_first-principles_2015}. 

The BPVE is much less studied in two-dimensional~(2D) materials~\cite{Zenkevich2014,cook_design_2015,lyanda-geller_polarization-dependent_2015}. Two-dimensional materials represent the ultimate scaling in thickness with mechanical, optical, and electronic properties that are unique relative to their bulk counterparts. For example, single-layer group-IV monochalcogenides \textrm{GeS}, \textrm{GeSe}, \textrm{SnS}, and \textrm{SnSe} are actively being investigated~\cite{li_single-layer_2013,singh_computational_2014,wang_substantial_2016,ramasamy_solution_2016,wu_intrinsic_2016,kamal_direct_2016,guo_thermoelectric_2016,xin_few-layer_2016,hanakata_polarization_2016} due to their band gaps and large carrier mobilities suitable for optoelectronics. Centrosymmetric 
in the bulk, the monochalcogenides lack inversion symmetry in single-layer form, allowing for the emergence of a 
spontaneous polarization and a BPVE. Although broken inversion symmetry is necessary for a nonzero shift current, the relationship between shift current and polarization at a given frequency is complex and depends on the degree of asymmetry and spatial localization of the valence and conduction states~\cite{young_first-principles_2012,*young_first_2012,*zheng_first-principles_2015,*young_first-principles_2015}. On the other hand, the shift-current spectrum integrated over frequency is clearly correlated to polarization, as shown in this Letter.
\begin{figure}[b]
\includegraphics{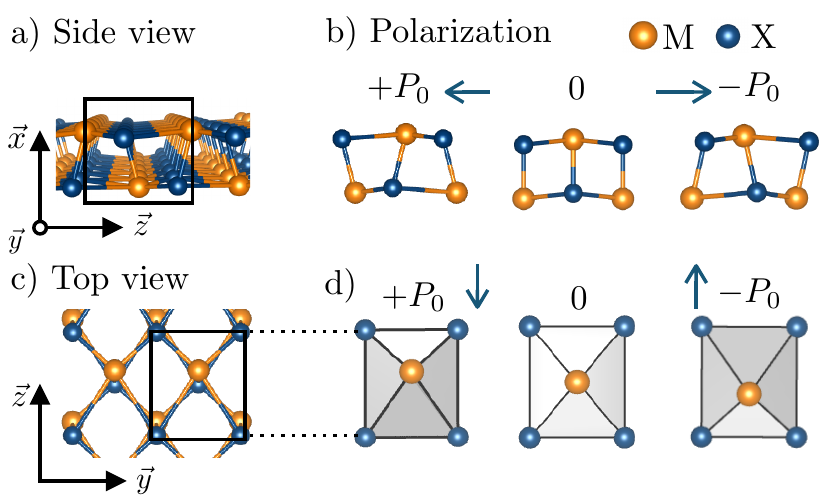}\\
\caption{(Color online) The crystal structure of single-layer group-IV monochalcogenides \textit{MX}, where 
\textit{M}=Ge, Sn, and \textit{X}=S, Se. In (a) we show the 3D view of the single-layer and in (b)-(d) the projections of the 
single layer crystal on the Cartesian axes. Structures with $0,\pm P_0$ polarizations are also shown in inset (b) 
and (d). A twofold rotation along $z$ (plus translations) determines the polarization axis, see text.}
\label{fig:geometry}
\end{figure}
In this Letter we use first principles density functional theory methods, supplemented by a tight-binding 
model, to predict and understand the spontaneous polarization and BPVE in single-layered monochalcogenides. 
In addition to confirming their established favorable band gaps and strong absorption~\cite{shi_anisotropic_2015,gomes_strongly_2016}, we demonstrate that the monochalcogenides exhibit a large in-plane shift current, up to $100$~$\mu$A/V$^2$. 
Using a Rice-Mele tight-binding model, we find that the integral of the frequency-dependent shift-current 
tensor is well correlated to the spontaneous polarization; and this integral is maximized at an optimal value 
of polarization.
%


\begin{table*}[]
\caption{Left: Ground state polarization of single-layer monochalcogenides. The 3D effective polarizations are 
for a layer thickness of $a= 2.6$~\AA. The energy barrier between the ground-states with opposite polarization 
calculated within DFT-PBE are also shown. Right: Direct~(D) and indirect~(I) band gaps calculated with DFT-PBE 
and optical gaps reported from $GW$-BSE calculations.}
\begin{tabular}{lccccccccccccc}
\toprule
\toprule
&\multicolumn{2}{c}{Polarization}
&\hspace*{0.15in}
& Energy 
&\hspace*{0.15in}
&\multicolumn{3}{c}{Supercell }
&\hspace*{0.15in}
& \multicolumn{4}{c}{Band gap (eV)}
\\
& ~~~~~~~2D~~~~~~ & ~~~~~~3D~~~~~~ && ~~~~barrier~~~~
 && \multicolumn{3}{c}{~~~~lattice vectors~(\AA)~~~~} 
&& \multicolumn{2}{c}{~DFT-PBE}~~~~~ & ~~~$GW$-BSE~~~ & ~~~~~Expt.$^a$\\
 & (nC/m) & (C/m$^2$) && (K) && $a$&~~~~~$b$&$c$ && D & I & D & D \\
\toprule
\textrm{GeS}  & 0.48 & 1.9 && 5563 && 15.0 & ~~~~~3.7 & 4.5 && 1.9 & 1.7&2.2~\cite{gomes_strongly_2016} & 1.6~\cite{ramasamy_solution_2016}\\
\textrm{GeSe} & 0.34 & 1.3 && 1180 && 15.0 &~~~~~ 4.0 & 4.3 && 1.2 & 1.2&1.6, 1.3~\cite{shi_anisotropic_2015,gomes_strongly_2016} & 1.2~\cite{ramasamy_solution_2016}\\
\textrm{SnS}  & 0.24 & 0.8 && 384  && 15.0 & ~~~~~4.1 & 4.3 && 1.5 & 1.4&\\                      
\textrm{SnSe} & 0.17 & 0.6 && 80   && 15.0 & ~~~~~4.3 & 4.4 && 0.9 & 0.9&1.4~\cite{shi_anisotropic_2015}\\                      
\bottomrule
\bottomrule
\end{tabular}\\
\begin{flushleft}{$^{a}$ Experimental optical gaps (Expt.) of few-layer chalcogenides are also shown for comparison.}
\end{flushleft}
\label{table:pol-gaps}
\end{table*}

{\it Structure, symmetries, and ab-initio methods}.-- Our DFT calculations are performed with the generalized 
gradient approximation including spin-orbit coupling. We use the ABINIT code~\cite{gonze_recent_2016}, 
with Gaussian pseudopotentials~\cite{krack_pseudopotentials_2005} and the Perdew-Burke-Ernzerhof~(PBE) 
functional~\cite{perdew_generalized_1996}. We fully relax atomic positions in supercells that include at 
least $10$~{\AA} of vacuum between layers. Our relaxed lattice parameters are shown in Table~\ref{table:pol-gaps} 
and agree with previous work~\cite{singh_computational_2014} (see details in Supplementary Material \cite{url_SM}, 
which includes Ref.~\cite{nastos_full_2007}).

Bulk monochalcogenide crystals \textit{MX} (\textit{M}=Ge, Sn and \textit{X}=S, Se) are orthorhombic with point 
group $mmm$ and space group $Pnma$ (No. $62$). They consist of van der Waals-bonded double layers of metal 
monochalchogenide atoms in an armchair arrangement. The space group of the bulk crystal contains eight symmetries 
including a center of inversion which prevents spontaneous electric polarization and the BPVE. Upon exfoliation, 
the resulting single ``double layer'' primitive cell has four atoms. In this work, the layers are chosen to be 
oriented perpendicular to the $x$ axis as shown in  Fig.~\ref{fig:geometry}(a). The single-layer structure has 
four symmetries, including a two-fold rotation with respect to $z$ (plus translation), $2[001] + (1/2,0,1/2)$, which 
determines the direction of the in-plane spontaneous polarization of the layer along the $z$ axis. In addition, the 
2D system has two mirror symmetries  with respect to $x$ and $y$, $m[100]+(1/2,1/2,1/2)$ and $m[010]+(0,1/2,0)$.
Hence its point group, which determines the nonzero components of the optical response tensors, is $mm2$. 

As a consequence of the mirror symmetries with respect to the $x$ and $y$ axis of a single monochalcogenide layer, 
all the cross-components terms of the imaginary part of the dielectric function, $\epsilon^{ab}_2$, vanish together 
with  the tensor components $xxx$, $xyy$, $xzz$, $yxx$, $yyy$, and $yzz$ of the shift current. Only seven components 
are symmetry allowed~\cite{Butcher1965}: $zxx$, $zyy$, $zzz$, $yyz$, and $xzx$, as well as components obtained by 
interchanging the last two indices. Symmetry, however, does not dictate the magnitude of the response in each direction, 
and consequently we compute the matrix elements below.

{\it Spontaneous polarization}.-- We calculate the spontaneous polarization of single-layer chalcogenides using 
the modern theory of polarization~\cite{vanderbilt_electric_1993,resta_electrical_2010}, as implemented in ABINIT. 
We first identify an adiabatic path between the ground state and a centrosymmetric geometry with, in this case, 
zero polarization. We parametrize the atomic displacements along a path between these geometries 
[Figs.~\ref{fig:geometry}(b) and ~\ref{fig:geometry}(d)] with $\lambda$ as $\v{R}^{i}(\lambda) = \v{R}^{i}_0 +\lambda (\v{R}^{i}_f-\v{R}^{i}_0)$, where $\v{R}^{i}_0$~($\v{R}^{i}_f$) is the  initial~(final) position of \textit{ith} atom in the 
centrosymmetric (noncentrosymmetric) structure. We calculate the minimum-energy path between the $\pm P_0$ 
configurations, as detailed in the Supplementary Material \cite{url_SM}. The minimal energy path is indistinguishable 
from the linear path used here. The polarization for various 2D monochalcogenides has also been theoretically 
studied recently~\cite{Gomes2015,Fei2015,mehboudi_structural_2016,fei_ferroelectricity_2016,hanakata_polarization_2016,wu_intrinsic_2016,mehboudi_two-dimensional_2016,wang_two-dimensional_2017}.  Our adiabatic polarization path, is shown schematically 
in Figs.~\ref{fig:geometry}(b) and Figs.~\ref{fig:geometry}(d). Table~\ref{table:pol-gaps} shows the computed spontaneous 
electric polarization per unit area, $P_0 a$, and an effective 3D polarization assuming an active single-layer 
thickness $a=2.6$~\AA. Interestingly, GeSe has a significantly higher effective 3D polarization, $1.9$ C/m$^2$, than 
most prototypical ferroelectrics, e.g., 0.0028 C/m$^2$ in \textrm{CaMn}$_7$O$_{12}$~\cite{Johnson2012}, 0.26 C/m$^2$ in 
\textrm{BaTiO}$_3$~\cite{Hippel1950,Shieh2009}, 0.37 C/m$^2$ in \textrm{KNbO}$_3$~\cite{Resta1993} and 0.9 C/m$^2$ in 
BiFeO$_3$~\cite{Neaton2005,Catalan2009}. The energy barriers, provided in Table~\ref{table:pol-gaps}, are much larger 
than room temperature. However, since reorientation under an applied electric field is often facilitated by domain 
wall motion, future experiments are necessary to conclusively demonstrate ferroelectric switching behavior.

{\it Optical absorption and shift current}.-- One notable feature of single-layer monochalcogenides is their promising 
band gap energies in the visible range~\cite{shi_anisotropic_2015,gomes_strongly_2016,ramasamy_solution_2016}.
In Table~\ref{table:pol-gaps}, we show the computed DFT-PBE gaps, which are a good estimate of the corresponding 
optical gaps. Although in principle \textit{ab initio} many-body perturbation theory~(MBPT) would be a more rigourous 
approach to optical gaps, in this case, we expect the PBE single-particle gaps to be indicative since excitonic 
effects, as large as 1 eV in \textrm{GeS}~\cite{gomes_strongly_2016}, can fortuitously cancel the well-known tendency 
of the PBE functional to underestimate the transport gap. In Sn-based materials, the PBE gaps are smaller by $\sim$0.5 eV 
than in prior MBPT  colculations (see Table~\ref{table:pol-gaps}), and hence the responses are redshifted and should 
be treated with more caution. In addition, since exciton formation can alter and even enhance shift current at exciton 
resonances~\cite{morimoto_topological_2016}, strong excitonic effects in monochalcogenides could lead to even larger 
shift currents.

\begin{figure}[]
\includegraphics{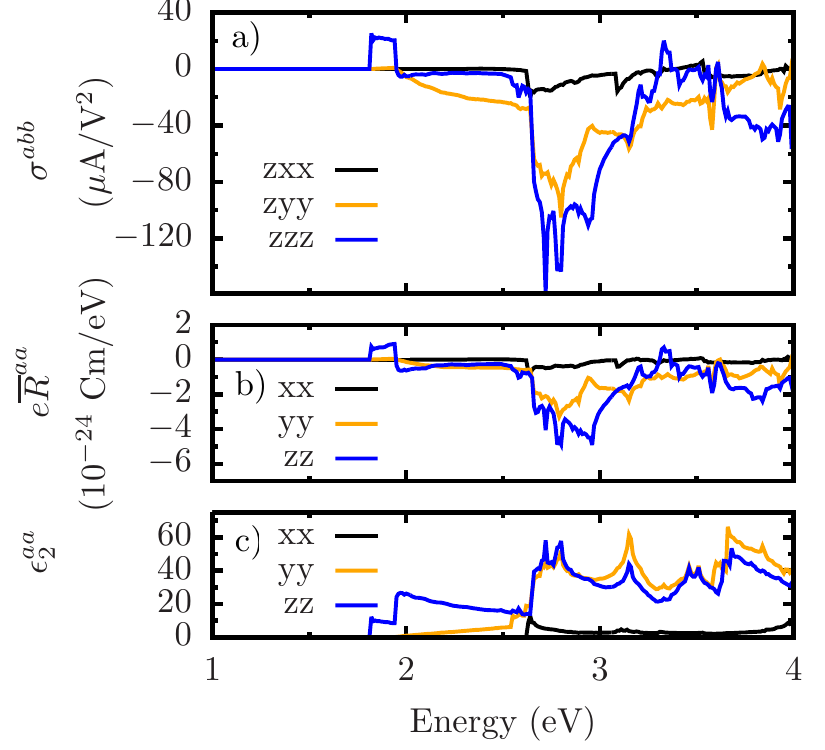}\\
\caption{(Color online)
Shift-current spectra (a), shift vector integrated over $\mathbf{k}$ (b), and linear absorption (c) of single-layer \textrm{GeS}.
The large in-plane shift-current response in the visible range is dominated by the shift vector and corresponds to the large absorption along $zz$ and $yy$.}
\label{fig:responses_main}
\end{figure}

We calculate the imaginary part of the the dielectric function, $\epsilon^{ab}_2$, within the independent particle 
approximation. As shown in Fig.~\ref{fig:responses_main}, the absorption is strong $\epsilon_2 \sim 50$, in the visible 
range of 1.5 to 3 eV, due to the direct or nearly direct band gap of these materials~\cite{cook_design_2015}. For comparison, we 
calculate the absorption coefficient $\alpha=\omega \epsilon_2$/c, with light frequency $\omega$ and speed of light $c$. 
For the 2D monochalcogenides of thickness $\sim 2.6$~\AA, $\alpha \sim 0.5$--$1.5\times10^6$cm$^{-1}$, and similar 
values were found for graphene and MoS$_2$ ($0.7$ and $1$--$1.5\times10^6$cm$^{-1}$, respectively)~\cite{bernardi_extraordinary_2013}. 
The $zz$ and $yy$ tensor components are larger than $xx$ due to the intrinsic crystal anisotropy, in agreement 
with previous work~\cite{shi_anisotropic_2015,ramasamy_solution_2016,gomes_strongly_2016}. In addition to the energy 
gaps and the large absorption in the visible range, single-layer monochalcogenides have a large shift-current response, as 
shown below.

The dc shift current is generated to second order in the electric field. Consider a monochromatic electric field of the 
form $E^b(t) = E^b(\omega)e^{i\omega t} + E^b(-\omega)e^{-i\omega t}$. The shift-current response can be expressed in 
terms of the third-rank tensor $\sigma^{abc}(0;\omega,-\omega)$ as, 
\begin{align}
J_{\textrm{shift}}^{a}(\omega) = 2 \sum_{bc}\sigma^{abc}(0;\omega,-\omega) E^{b}(\omega)E^{c}(-\omega).
\label{eq:def_shift_current}
\end{align}
The shift-current tensor is given by~\cite{sipe_second-order_2000}
\begin{align}
\sigma^{abc}(0;\omega,-\omega) &= -\frac{i \pi e^3}{2 \hbar^2} \int \frac{d \v{k}}{8\pi^3} \sum_{nm} f_{nm} \big( r^b_{mn} r^c_{nm;a} \nonumber\\
&~~~~~~~+ r^c_{mn} r^b_{nm;a} \big) \delta(\omega_{mn}-\omega),
\label{eq:sigma_main}
\end{align}
where $r^{a}_{mn}$ are velocity matrix elements. The $r^{a}_{mn;b}$ are generalized derivatives, 
defined as $r^{a}_{mn;b} = \partial r^a_{mn}/ \partial {k^b} - i (A^{b}_{nn}- A^{b}_{mm}) r_{nm}^{a}$, where 
$A^{a}_{nm}$ are the Berry connections, with the $a$ and $b$ indices denoting Cartesian directions. We define the Fermi-Dirac 
occupation numbers $f_{nm}= f_n - f_m$, and the band energy differences as $\hbar \omega_{nm} = \hbar\omega_n - \hbar \omega_m$. 
For linearly polarized incident light, $b=c$, and the integrand in Eq.~(\ref{eq:sigma_main}) is proportional to the shift 
``vector'' $R_{nm}^{ab}$~\cite{sipe_second-order_2000}, defined as $(1/2) \mathrm{Im}[r^b_{nm}r^b_{mn;a} - r^b_{mn} r^b_{mn;a}] |r^b_{nm}|^{-2}$.

Figure~\ref{fig:responses_main} shows the calculated effective shift-current spectra for GeS parallel and perpendicular 
to the polarization axis. (The shift-current spectra for GeSe, SnS, and SnSe have similar features, see Supplementary 
Material \cite{url_SM}.) We report the responses assuming an active single-layer thickness of 
$a=2.6$~\AA~
\footnote{Here we show that a single-layer monochalcogenide has a large shift current response, moreover, in the 
Supplementary Material \cite{url_SM} we show that the response of a thick 3D array of slabs, characterized by the 
Glass coefficient~\cite{Glass1974}, is also large.}
~We find a broad maximum of the order of 100 
$\mu$A/V$^2$ which, importantly, occurs in the visible range ($1.5 - 3.3$ eV). The in-plane components, $zzz$ and 
$zyy$, are larger than the out-of-plane component $zxx$, consistent with the large absorption along $zz$ and $yy$. 
We compare this response with that of prototypical ferroelectric materials in the same frequency range, e.g., 
$0.05$~$\mu$A/V$^2$ in \textrm{BiFeO}$_3$~\cite{young_first-principles_2012}, and $5$ ~$\mu$A/V$^2$ in 
\textrm{BaTiO}$_3$~\cite{young_first_2012}, which are much smaller. Additionally, $0.5$~$\mu$A/V$^2$ is reported for 
hybrid halide perovskites~\cite{zheng_first-principles_2015} and \textrm{NaAsSe}$_2$~\cite{Brehm2014}, and  250 
$\mu$A/V$^2$ (= 400 mA/W) is found for state-of-the-art Si-based solar cells~\cite{Pagliaro2008} 
(see the Supplementary Material \cite{url_SM} for details on the conversion between shift-current and A/W units). 
The BPVE for  2D monochalcogenides is therefore quite large.

The absorption and shift-current spectra are related by the velocity matrix elements $r_{nm}$ entering Eq.~(\ref{eq:sigma_main}), 
explaining why peaks in $\epsilon_2$ tend to correspond to peaks in the shift current spectra. To explore the relationship 
between $\epsilon_2^{bb}$ and $\sigma^{abb}$, in Fig.~\ref{fig:responses_main}(b), we plot the shift vector integrated over the 
Brillouin zone~(BZ)~\cite{young_first-principles_2012,*young_first_2012,*zheng_first-principles_2015,*young_first-principles_2015,fregoso_quantitative_2017},
\begin{align}
e\overline{R}^{ab}(\omega) = e\Omega  \int \frac{d\mathbf{k}}{8\pi^3} \sum_{nm} f_{nm} R^{ab}_{nm} \delta(\omega_{nm}-\omega), 
\label{eq:int-Rcv}
\end{align}
where $\Omega$ is the volume of the unit cell. For the monochalcogenides, $e\overline{R}^{bb}(\omega)$ contains 
most features of $\sigma^{abb}(\omega)$, and hence 
dominates the shift-current response. Following the analysis of Ref.~\cite{fregoso_quantitative_2017}, we interpret 
$e\overline{R}^{ab}$ as a collective shift of polarization upon excitation, due to transitions from valence to conduction 
states with a distinct center of mass~\footnote{The integral of the shift vector over $\mathbf{k}$, $\overline{R}$ has 
been studied for complex ferroelectric materials, such as BiTiO$_3$ and PbTiO$_3$ in Ref.~\onlinecite{young_first_2012}.}.	 
Thus, wave function Berry phases play a fundamental role in the BPVE, which we further explore next.

\begin{figure}[]
\includegraphics{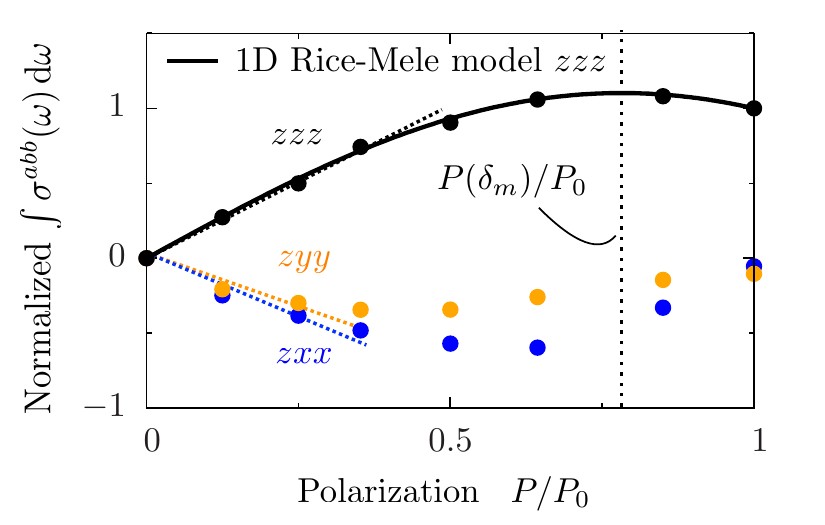}\\
\caption{(Color online) Nonmonotonic dependence of the integral of the shift-current tensor vs electric polarization 
for \textrm{GeS}; the integral is normalized by $-3\times10^{10}$~As$^{-1}$V$^{-2}$, its value at the ground state 
with polarization $P_0=1.9$ C/m$^{2}$. The tensor components $zzz$, $zyy$, and $zxx$ are shown in black, green, and 
blue points, respectively. For small $P$ the integral is directly proportional to polarization (dashed lines), but 
it is nonmonotonic for large $P$. The integral reaches it maxima at $P(\delta_m)$, which is close to $P_0$.}
\label{fig:polarization_main}
\end{figure}

{\it Polarization and shift current}.-- To understand the relation between $\sigma^{abb}$ and polarization $P$, 
consider a short-circuit bulk ferroelectric illuminated by unpolarized light with a flat broad spectrum.
The short-circuit current in the $z$ direction,
$I_{\mathrm{sc}} = A E_0^2 \int d\omega (\sigma^{zyy}(0;\omega,-\omega) +\sigma^{zzz}(0;\omega,-\omega))$, 
is proportional to the integral of the shift current tensor. Here, the cross sectional area is $A$ and the 
amplitude of the electric field is $E_0$. To first nonvanishing order in $\lambda$, both the polarization 
and the integral are linear in $\lambda$, $P(\lambda) = \partial_\lambda P(0)~\lambda + \cdots$,   
$\int \sigma^{abb}(\lambda) = (\int \partial_\lambda\sigma^{abb}(0))~\lambda + \cdots$, 
and hence proportional to each other. 

In Fig.~\ref{fig:polarization_main} we show the integral of the shift-current tensor over the frequency range 
(up to $6$~eV) for GeS as a function of polarization along the adiabatic path of Fig.~\ref{fig:geometry}(b). 
For small polarization, the integral grows linearly with polarization, as expected.
However, for larger polarization there is nonmonotonic behavior which we explain below with a tight-binding model. 
Notice that without the integral, the expansion coefficients become frequency dependent and the current could 
increase or decrease with polarization with no general relationship.

\textit{The Rice-Mele tight-binding model}.-- As mentioned previously, the monochalcogenide layer has a 
$2[001] + (1/2,0,1/2)$ symmetry that transforms the upper three atoms in Fig.~\ref{fig:geometry}(b) onto the lower 
three. This suggests there is an effective one-dimensional description of the armchair structure in the $z$ direction, 
and in fact, as we show below, the trends in the integral of the shift-current tensor along $z$ are captured by a 
simple model Rice-Mele (RM) model~\cite{Vanderbilt1993,Onoda2004}. The RM Hamiltonian is 
\begin{align}
H= \sum_{i} \bigg[ (\frac{t}{2} + (-1)^{i}\frac{\delta}{2}) (c_{i}^{\dagger} c_{i+1} + h.c.) + (-1)^{i} \Delta c^{\dagger}_{i} c_{i} \bigg],
\end{align}
where $\delta$ parametrizes the structural distortion relative to the centrosymmetric structure, $\Delta$ the 
staggered on-site potential, and $c_{i}^{\dagger}$ is the creation operator for electrons at site $i$. Inversion 
symmetry is broken when both $\Delta\neq 0$ and $\delta\neq 0$, and preserved otherwise. For this two-band model 
Eq.~(\ref{eq:sigma_main}) gives (see Supplementary Material \cite{url_SM} for more details), 
\begin{align}
\int d \, \omega~\sigma^{zzz}(0;\omega,-\omega) &= e^3 \int d k \, \frac{ |v_{cv}|^2  R_{cv}}{4 E^2}, 
\label{eq:int_sigma_zzz}
\end{align}
where $R_{cv} =\partial\phi_{cv}/\partial k + A_{cc} -A_{vv}$ is the shift vector and is gauge invariant. 
$A_{nm} = i \langle u_n| \partial_k |u_m\rangle$ are the Berry connections, and $\phi^b_{nm}$ is
defined by $r^b_{nm}=|r^b_{nm}|e^{-i \phi^b_{nm}}$, where $E(k)$ is the band dispersion and $v_{cv}$ is the 
matrix element of the velocity operator. The Rice-Mele model allows for a complete analytic solution for 
the optical response and these results will be presented elsewhere~\cite{Fregoso}. The polarization is 
$P(\delta) = (e/2\pi)\int dk ~[A_{vv}(k,\delta)-A_{vv}(k,0)]$. The model has two independent parameters 
$\delta$ and $\Delta$. To make contact with the monochalcogenides, we set $t=1$, and $\delta$ and $\Delta$ 
are related by the energy gap, $2\sqrt{\delta^2_0 + \Delta^2_0}=1.9$~eV (for GeS). Choosing parameters 
$(t,[0,\delta_0],\Delta_0)=(1,[0,-0.87],0.4)$~eV fits the $zzz$ \textit{ab initio} integral for \textrm{GeS} well 
and corresponds to a gap of $1.9$~eV. The RM model polarization is $P_0=P(\delta_0)$. 

As the RM model is a good description of the integral of the shift-current tensor in monochalcogenides, 
we now explore the relation between polarization and shift current within this model. The integral of the shift 
current tensor in Eq.~(\ref{eq:int_sigma_zzz}) is determined by the competition between the shift vector and 
the velocity matrix elements $\hbar^{2}|v_{cv}|^2/4E^2 \equiv |r_{cv}|^2$. These in turn are controlled by 
$\delta$ and $\Delta$, which have opposing tendencies: whereas increasing $\Delta$ tends to localize charge 
at lattice sites, increasing dimerization $\delta$ moves the center of charge away from them (leading to an 
increase in polarization).  We find that for $\delta \ll \Delta$, $R_{cv}$ is sharply peaked at $k=0$ but 
$|v_{cv}|^2/4E^2$ peaks at $\pi/c$, and hence the integral is small. As $\delta$ increases, $R_{cv}$ and 
$|v_{cv}|^2 /4E^2$ broaden, and the integral increases; the integral reaches a maximum at an optimum value, 
$\pm \delta_m$ which to lowest order in $\Delta$ is
\begin{align}
\delta_m= \Delta + O(\Delta^3 \mathrm{log} \Delta),
\end{align}
where the polarization takes the value $P(\delta_m)$. For GeS, GeSe, and SnS, $\delta_0$ ($-0.9$, $-0.5$, and $-0.6$ respectively) 
is relatively close to the optimal $\delta_m$ values of $-0.5$, $-0.5$, and $-0.6$, respectively, whereas for 
SnSe $\delta_m$ ($0.4$) is farthest from $\delta_0$ ($-0.2$), see the Supplementary Material \cite{url_SM}.
Therefore, consistent with Refs.~\onlinecite{cook_design_2015} and~\onlinecite{young_first-principles_2012,*young_first_2012,*zheng_first-principles_2015,*young_first-principles_2015}, the large shift current in these monochalcogenides results from two competing factors, a large shift vector and large velocity matrix elements (linear absorption strength), both of which can be modulated with polarization (and therefore composition, structure, and external electric field).

{\it Discussion and conclusions}.-- We have calculated the shift-current response and spontaneous electric polarization 
of a family single-layer monochalcogenides, \textrm{MX}, where \textit{M}=Ge, Sn, and \textit{X}=S, Se. We find a large 
shift current and a large polarization compared with prototypical ferroelectric materials. The fact that the maximum current 
occurs in the visible range highlights the potential of these materials for optoelectronic applications. Further, the 
large spontaneous polarization can serve as a knob to engineer the photoresponse. 
The integral of the shift-current tensor over frequency is clearly dependent on polarization and by means of a RM model, we find an optimal value of polarization where the current is maximum.

\begin{acknowledgments}
We thank J. Sipe, F. de Juan, S. Barraza-Lopez, S. Coh, R. A. Muniz, and S. E. Reyes-Lillo for useful discussions.
This work is supported by the U.S. Department of Energy, Director, Office of Science, Office of Basic Energy Sciences, Materials Sciences and Engineering Division, under Contract No. DE-AC02-05CH11231, through the Theory FWP (KC2301) at Lawrence 
Berkeley National Laboratory (LBNL). B.M.F. acknowledges support from AFOSR MURI, Conacyt, and NERSC Contract No.
DE-AC02-05CH11231. This work is also supported by the Molecular Foundry through the DOE, Office of Basic Energy Sciences 
under the same contract number. T.M. acknowledges support from the Gordon and Betty Moore Foundation's EPiQS Initiative 
Theory Center Grant. J.E.M. acknowledges Laboratory Directed Research and Development funding from LBNL Contract No. 
DEAC02-05CH11231. B.S.M. acknowledges partial support from CONACYT-Mexico GoGa No. 153930. We acknowledge the use of computational resources at the NERSC.

\noindent B.M.F. and 	T.R. contributed equally to this work.
\end{acknowledgments}
 

%


\appendix

\section{Supplementary information: Large bulk photovoltaic effect and spontaneous polarization of single-layer monochalcogenides}
In this Supplementary Information we present details of our electric polarization and shift current calculation for \textrm{GeS}, \textrm{GeSe}, \textrm{SnS} and \textrm{SnSe}. 
We also provide details of the shift current and polarization in the Rice-Mele model. 

\section{Numerical details}
\label{sect:tec-details}
Our density functional theory~(DFT) calculations are done with the \textrm{ABINIT} code~\cite{gonze_recent_2016} 
and within the Perdew-Burke-Ernzerhof functional~(PBE)~\cite{perdew_generalized_1996}. We use Hartwigsen-Goedecker-Hutter 
norm conserving pseudopotentials available in the \textrm{ABINIT} web site~\cite{abinit_site}. We use an energy 
cut-off of $40$ hartrees to expand the plane-wave basis set. To model the slabs we use supercells with $15$~\AA\ 
along the non-periodic direction, which corresponds to $>10$~\AA\ of vacuum. To calculate $\sigma$ we include 
$20$~valence and $30$~conduction bands, which accounts for all allowed transitions in the low energy range; up 
to~$6$~eV; we use a mesh of $70 \times 70$~k-points along the periodic slab directions and integrate using a 
tetrahedron method. Worth mentioning that the optical-response and spontaneous-polarization magnitude depend 
on the calculation volume, therefore we renormalize our results to the atomic slab widths (removing the vacuum) of 
$2.56$, $2.61$, $2.84$ and $2.73$~\AA\ for \textrm{GeS}, \textrm{GeSe}, \textrm{SnS} and \textrm{SnSe} respectively.
String method calculations were done with the \textrm{ABINIT} code as explained in Sect. 3.2 of 
Ref.~\onlinecite{gonze_recent_2016}, where the minimal energy path between two points was found using 50 images and a 
tolerance on the mean total energy for images of $2 \times 10^{-6}$~hartrees. In the next, we provide more details on 
our calculations of spontaneous polarization, the ferroelectric energy barrier and optical responses of the monochalcogenides.

\section{Spontaneous Polarization}
\label{sect:polarization}
\begin{figure}[]
\includegraphics[width=0.48\textwidth]{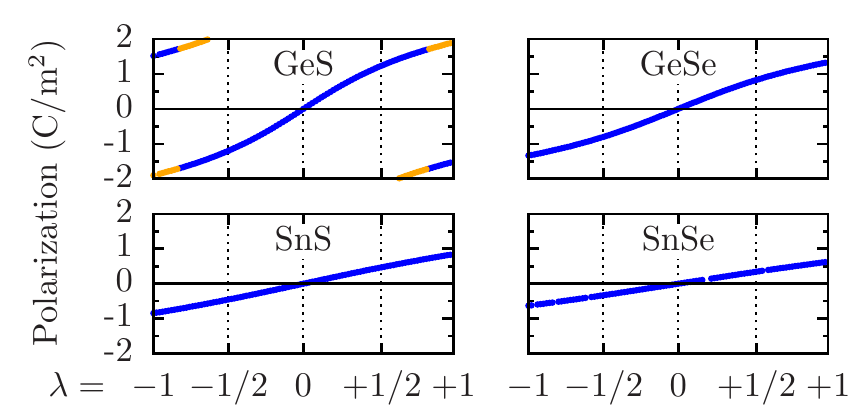}
\caption{Polarization along adiabatic path parametrized by $\lambda$. The gap does not closes along the 
path and takes the system from polarization $-P_0$~($\lambda=-1$) to 0 ($\lambda=0$) to $+P_0$~($\lambda=+1$). 
The calculated polarization (blue dots) is shifted by the twice polarization 
quantum~($Q=1.7$~ C/m$^2$) for \textrm{GeS} (orange dots).}
\label{fig:polarization}
\end{figure}
We calculate the spontaneous polarization as implemented in ABINIT using
\begin{equation}
P^{a}(\lambda) = \sum_{i} \frac{eZ^{i}r^{a}_{i}(\lambda)}{\Omega}-\mathrm{i} e \sum_v \int_{\textrm{BZ}}\frac{ d\mathbf{k}}{(2\pi)^3} \langle u_{v}^{\lambda} | \nabla^{a} u_{v}^{\lambda} \rangle,
\end{equation}
where $u_{v}^{\lambda}$ are Bloch wave functions, $Z^{i}$ is the atomic number of the $ith$ atom, $\Omega$ 
is the simulation volume. $\lambda$ parametrizes an adiabatic path from a centrosymmetric configuration to 
the ground-state configuration. The polarization is defined as the difference between the polarization of 
two smoothly connected atomic structures: $\v{R}^{i}_0$ with inversion symmetry (i.e., zero of polarization) 
and $\v{R}^{i}_f (\lambda=1)$, where $\v{R}^{i}_f = \v{R}^i_0 + \lambda (\v{R}^{i}_f - \v{R}^{i}_0)$. The geometry 
of the $\v{R}^{i}_f$ and $\v{R}^{i}_0$ points used in this work are shown in Table~\ref{table:coords}. The polarization 
calculated at small steps of $\lambda$ for the different crystals is shown in Fig.~\ref{fig:polarization}. Since all 
points are connected smoothly, the polarization is well-defined and can be calculated as the difference between 
the polarization at $\v{R}^{i}_0$ and $\v{R}^{i}_f$. The resulting spontaneous polarization values of the different 
crystals are tabulated in Table. 1 of the main manuscript.

\section{Minimum energy path}
\label{sect:string}
\begin{figure}[]
\includegraphics{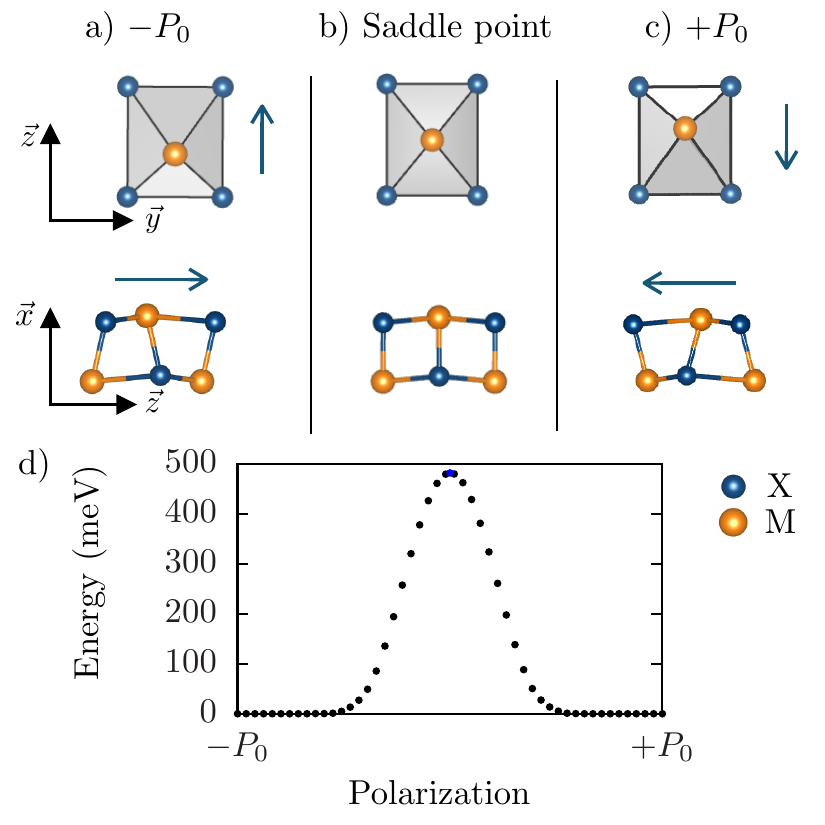}
\caption{Ferroelectric energy barrier of \textrm{GeS} calculated using the string method.
Insets a-c: ground state configurations with polarization $\pm P_0$ and saddle point.
Inset d shows the calculated energy barrier, with the saddle point in blue.}
\label{fig:string}
\end{figure}
Here we detail how we evaluate the ferroelectric energy barrier between the two ground-states with inverse polarization $\pm P_0$.
Here we assume the NVT ensemble, i.e. we keep the cell-dimensions constant.
We first calculate a path of fifty equidistant-potential points connecting the two frontier points using the string method.
Note that with this grid of points we cannot evaluate the polarity of the central point due to numerical remaining errors.
Hence, to obtain the precise coordinates of the saddle point, for \textrm{GeS} we calculate a finer grid of thirty equidistant-potential points in between the two central points of the fifty-point initial path.
In Fig.~\ref{fig:string} we show the energy barrier and initial, central and final configurations along the trajectory for \textrm{GeS}.
The resulting coordinates of the highest-energy saddle point (Fig.~\ref{fig:string}b) are 
\begin{align}
\textrm{Atom coordinates (\AA):} \nonumber\\
\begin{tabular}{lccc}
\textrm{Ge} & 1.54 & 0.91 & 0.00\\
\textrm{Ge} & 3.84 & 2.74 & 2.24\\
\textrm{S}  & 3.86 & 0.91 & 0.00\\
\textrm{S}  & 1.51 & 2.74 & 2.24 \nonumber\\
\end{tabular}
\end{align}
which resemble those of the ideal $R_0$ point (see Fig.1 of the main text and coordinates in Table~\ref{table:coords}); 
while the $x$ Cartesian component is slightly modified by $\sim 0.1$~bohr, the $y$ and $z$ Cartesian components are 
identical, implying zero-polarization along $z$. In terms of energetics, the saddle-point total energy is $\sim 70$~meV 
lower than that of the $R_0$ point.

\section{Electronic bandstructures}
\label{sect:bandstructures}
\begin{figure}[t]
\includegraphics[width=.45\textwidth]{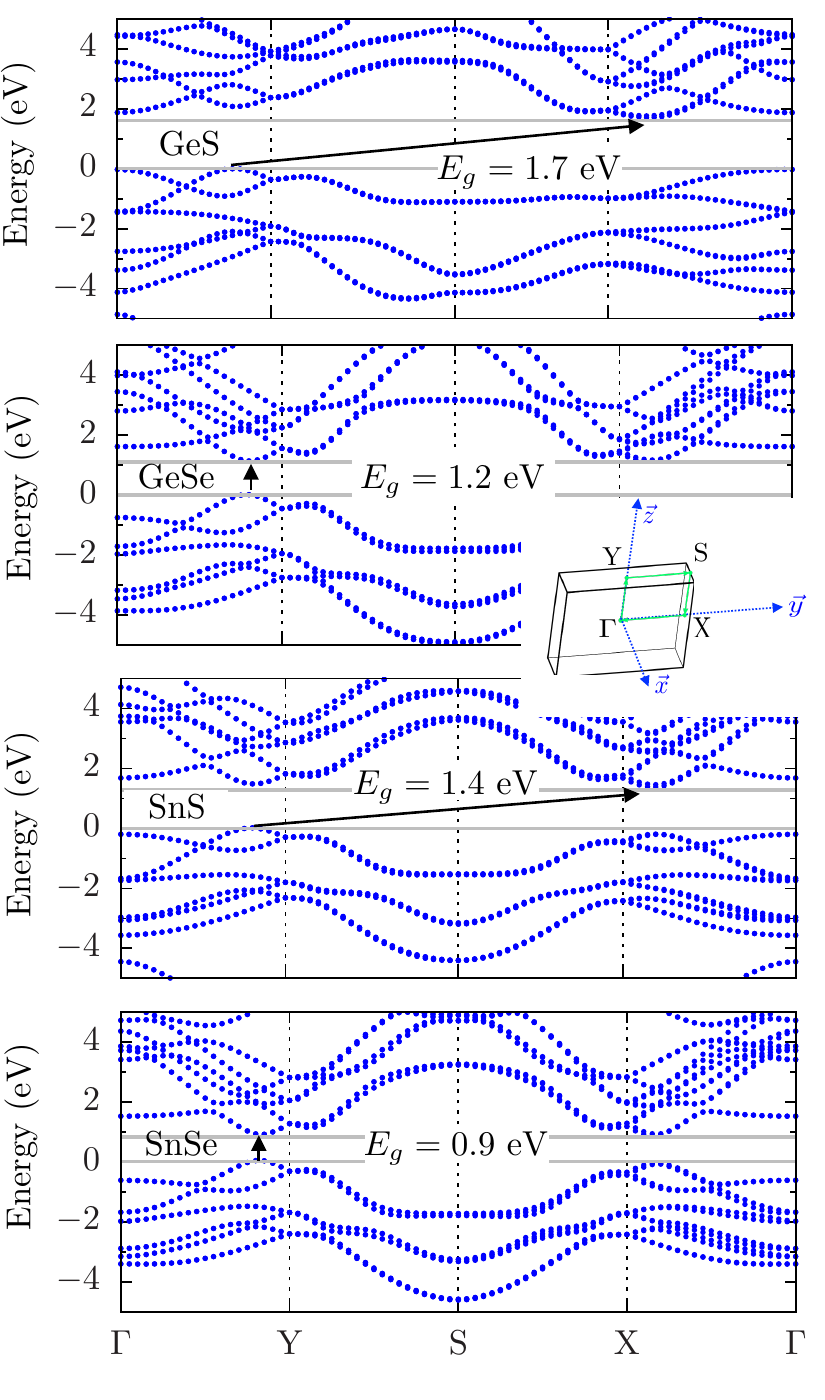}
\caption{Electronic bandstructure of group-IV single-layer monochalcogenides, calculated within DFT-PBE. We choose a 
k-point path along the Brillouin zone, shown at the bottom.}
\label{fig:bandstructure}
\end{figure}

The electronic bandstructures calculated within DFT are shown in Fig.~\ref{fig:bandstructure}, these agree with 
previous works~\cite{singh_computational_2014}. For each material the fundamental gap~$E_g$ is indicated in the 
figure with an arrow.

\section{Linear and shift-current spectra}
\label{sect:shift-current}
\begin{figure}[]
\includegraphics[]{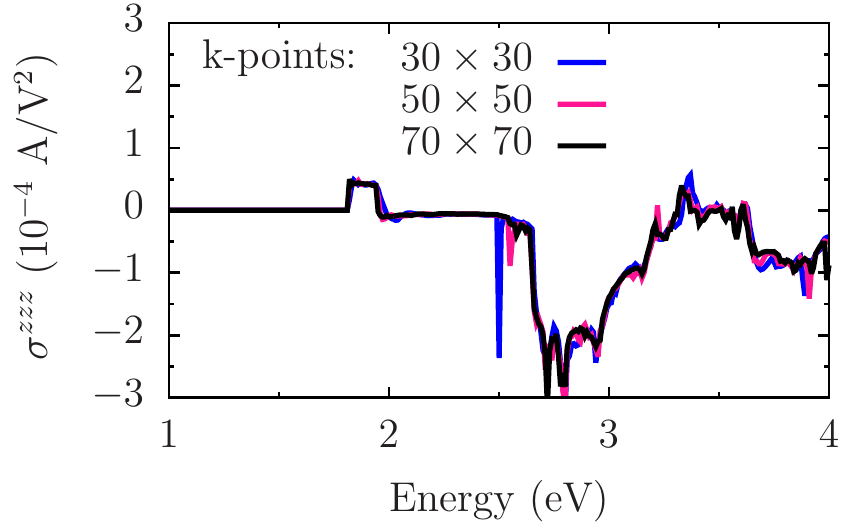}
\caption{Convergence of shift-current tensor for \textrm{GeS} with respect to the $\mathbf{k}$-mesh size.} 
\label{fig:conv}
\end{figure}
We calculate the linear absorption within the independent-particle approximation,
\begin{align}
 \epsilon^{ab}_2(\omega)&= \delta_{ab}-\frac{e^2 \pi}{\epsilon_0\hbar} \int \frac{d \v{k}}{(2\pi)^3} \sum_{nm} f_{nm} r^a_{nm}r^b_{mn} \delta(\omega_{mn} - \omega),
\label{eq:epsilon2}
\end{align}
which provides a RPA frequency dependent optical absorption spectrum. Here 
$r^a_{nm} = A^a_{nm} = i\langle u_n|\nabla^a|u_m\rangle$  ($n\neq m$) are the velocity matrix elements and 
$A^{a}_{nm}$ the usual Berry connections, with the $a$ and $b$ indices indicating one of the three Cartesian 
directions. We define the occupations numbers $f_{nm}= f_n - f_m$,  and the band energy differences as 
$\hbar \omega_{nm} = \hbar\omega n - \hbar \omega_m$. Here and in what follows we assume zero temperature. 

The shift-current response is calculated from Eq. (2) of the manuscript,
\begin{align}
\sigma^{abc}(0;\omega,-\omega) &= -\frac{i \pi e^3}{2 \hbar^2} \int \frac{d \v{k}}{8\pi^3} \sum_{nm} f_{nm} \big( r^b_{mn} r^c_{nm;a} \nonumber\\
&~~~~~~~+ r^c_{mn} r^b_{nm;a} \big) \delta(\omega_{mn}-\omega).
\label{eq:sigma}
\end{align}
First, we calculate the wavefunctions $\psi_m$ and eigenvalues using \textrm{ABINIT}.
The velocity matrix elements are computed with a plane-wave expansion,
\begin{align}
\psi_n(\mathbf{k};\mathbf{r})=
\sum_\mathbf{G} C_{m\mathbf{k}} (\mathbf{G}) 
e^{i(\mathbf{k}+\mathbf{G})\cdot \mathbf{r}},
\end{align}
such as, 
\begin{align}
r_{mn}(\mathbf{k}) = \hbar \sum_\mathbf{G}
C^*_{m\mathbf{k}}(\mathbf{G})
C_{n\mathbf{k}}(\mathbf{G})
(\mathbf{k}+\mathbf{G})
\end{align}
where $C_{m\mathbf{k}}(\mathbf{G})$ are expansion coefficients of the plane-wave $\mathbf{G}$ vectors. Subsequently, 
the generalized derivatives $r_{nm;b}$ are calculated from velocity matrix elements as in 
Ref.~\cite{sipe_second-order_2000} (see Section VIII). We accelerate convergence on $\mathbf{k}$ points by using a 
tetrahedrum-integration method~\cite{nastos_full_2007}. The optical responses are calculated with the \textrm{Tiniba} 
code, see further details in Ref.~\onlinecite{tiniba}.

We noticed that shift-current response converges relatively slow with respect to the number of k-points.
As shown in Fig.~\ref{fig:conv}, for \textrm{GeS} we require a dense mesh of $70 \times 70$~k-points on 
the slab plane to converge, removing sudden jumps (e.g., see blue and pink lines close 2.5 eV) in the response.

\begin{figure*}[]
\includegraphics[width=1.0\textwidth]{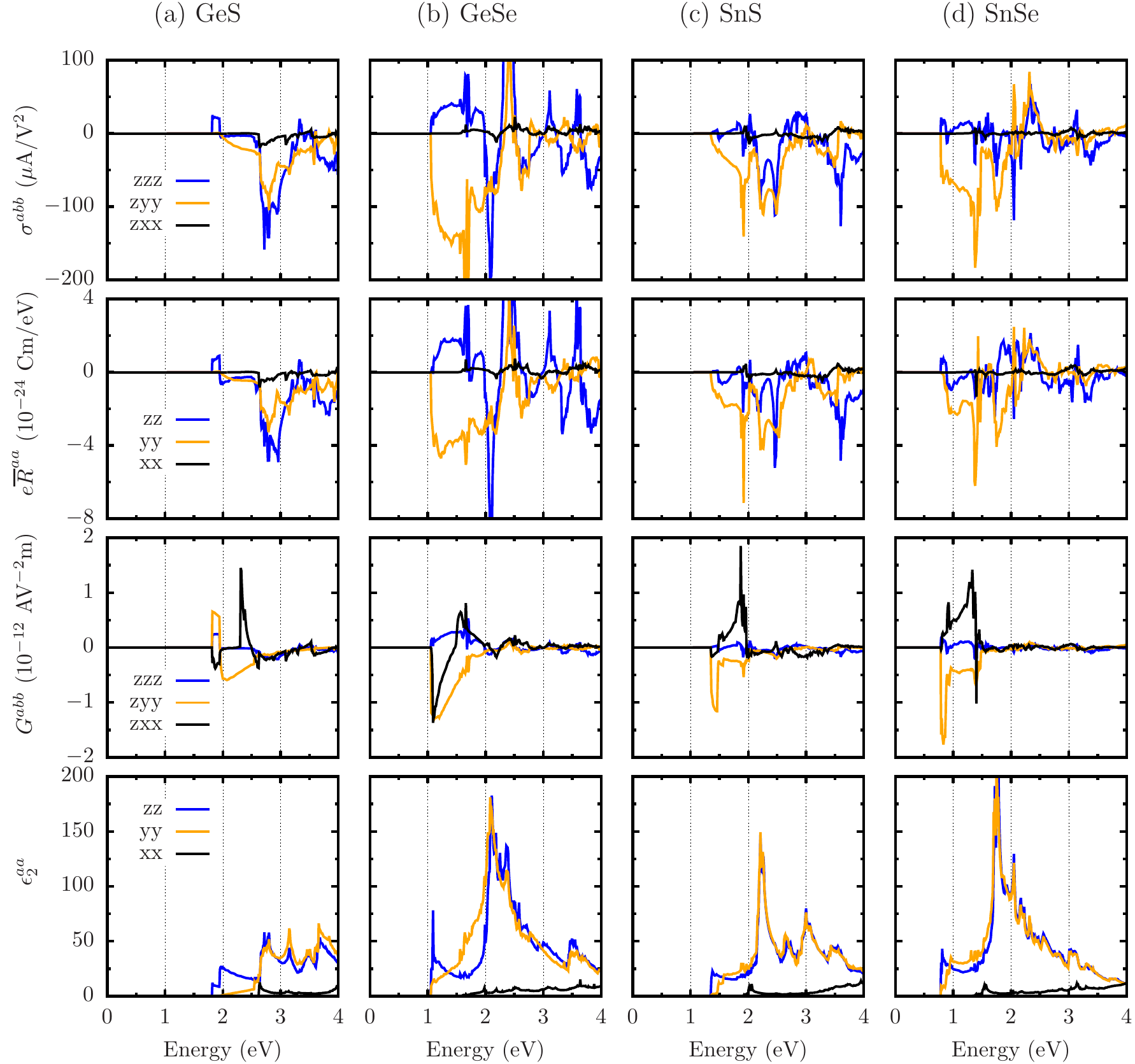}
\caption{Shift current spectra, shift vector,  linear response and Glass coefficient of single-layer monochalcogenides. 
The large shift current, Glass coefficient and linear absorption ($\epsilon_2\sim 100$) induced in the visible range of 
frequencies indicate their potential for optoelectronic applications.}
\label{fig:responses}
\end{figure*}

In Fig.~\ref{fig:responses}, we show our calculated linear and shift-current responses of the layered monochalcogenide 
materials studied in this work. As mentioned in the main text, the responses are related by the matrix elements 
$|r_{nm}|^2$ entering in Eqs.~\ref{eq:epsilon2} and \ref{eq:sigma}. Therefore, peaks in $\epsilon_2$ tend to correspond 
to peaks in $\sigma$, and the shift-current responses along $zzz$ and $zyy$ are larger than for $zxx$ in consistency with 
the large in-plane linear response.

In Fig.~\ref{fig:responses} we also show the shift vector integrated over the Brillouin zone $e\overline{R}^{aa}$ 
for the monochalcogenides. The integral of the shift vector has been previously studied for complex oxides, such 
as BiTiO$_3$ and PbTiO$_3$ in Ref.~\onlinecite{young_first_2012} where no obvious relation was found between the shift 
vector and the shift current spectra, instead the relation was found to depend on the degree of localization between 
initial and final states in the optical transitions. On the contrary, in the single-layer monochalcogenides, with 
delocalized $p$-type valence and conduction states, the relation between the shift current spectra and $e\overline{R}^{aa}$ is pronounced, e.g., the integrated shift vector shows all of the features in $\sigma$. This shows that the shift vector 
clearly drives the shift current response of the monochalcogenides, in other words  wavefunction Berry phases play a 
clear role in the shift current spectra. Following the analysis of Ref.~\cite{fregoso_quantitative_2017} we provide 
an interpretation of the shift vector in the main text, which complements previous analysis of the shift vector given 
in Refs.~\cite{sipe_second-order_2000,young_first-principles_2012,young_first_2012,zheng_first-principles_2015,young_first-principles_2015,fregoso_quantitative_2017}. 

In Fig.~\ref{fig:responses} we also show the Glass coefficient,
\begin{equation}
G^{abb}(\omega)=\sigma^{abb}(\omega)/\alpha^{bb}(\omega);
\end{equation}
where $\alpha^{bb}=\epsilon^{bb} \omega$ is the absorption coefficient~\cite{Glass1974}. The Glass coefficient 
defines the photocurrent of a thick sample (a thick 3D array of slabs in this case) accounting for the incident-light 
penetration depth ($\propto 1/\alpha$). The calculated $G^{abb}$ for the single-layer monochalcogenides, of the order 
of $\sim 10^{-10}$ cmAV$^{-2}$, is relatively large compared to that in prototypical ferroelectric materials, e.g., 
$\sim 10^{-11}$ cmAV$^{-2}$ in BiFeO$_3$~\cite{young_first-principles_2012} and $1.3$~$10^{-11}$ cmAV$^{-2}$ in 
BiTiO$_3$~\cite{young_first_2012}.

\subsection{Shift current equivalent in SI}
In this section we detail how to convert the shift-current tensor $\sigma$ to current density generated 
per unit intensity of light $\kappa$ in units of (A/m$^2$)/(W/m$^2$), and viceversa. We assume a linearly 
polarized light source of unitary intensity, $I=c \epsilon_0 E^2/2$, where $c$ is the speed of light and 
$\epsilon_0$ is the vacuum permitivity. The shift-current respose $J_\textrm{shift}$ is expressed in terms 
of the shift-current tensor $\sigma$ in Eq.~1 of the main text. Hence the current density is, 
$\kappa \equiv J_\textrm{shift}/I=  4\sigma /(c \epsilon_0)$. For example, given a $\kappa = 400$ mA/W 
(at given frequency), say in units of (A/m$^2$)/(W/$m^2$) = A/W, we can convert $\kappa$ into equivalent 
units of shift-current tensor. 

\section{Shift current and polarization in the Rice-Mele model}
\label{sect:rice-mele}
In this section we provide some details of the derivation of the shift current for the one dimensional (1D) 
Rice-Mele (RM) model. Starting from Eq. (2) in the main text we set $b=c$ (linear polarization) and using 
$r^b_{nm}=|r_{nm}^b| e^{-i \phi^b_{nm}}$ and $r^{a}_{mn;b} = \partial r^a_{mn}/ \partial {k^b} - i (A^{b}_{nn}- A^{b}_{mm}) r_{nm}^{a}$,
Eq.~(2) takes the familiar form
\begin{align}
\sigma^{abb}(0;\omega,-\omega) &=  - \frac{\pi e^3}{\hbar^2} \times 
 \nonumber\\ 
& \int \frac{d^d k}{(2\pi)^{d}} \sum_{nm} f_{nm} |r^{b}_{nm}|^2 R^{a,b}_{nm}
\delta(\omega_{nm} -\omega),
\label{eq:shift_current_shift_vector}
\end{align}
in terms of the shift vector is $R^{a,b}_{nm} =\partial \phi^{b}_{nm}/\partial k^a + A^{a}_{nn} -A^{a}_{mm}$.
For a two-band model in 1D, $n,m$ take two values $c,v$, and hence we obtain Eq.~6 of the main text. In the 
derivation we used $r^{a}_{nm} = \langle u_n|v^{a}|u_m \rangle/i \omega_{nm}$ that is valid for non-degenerate 
bands. Now we apply this results to the RM Hamiltonian Eq. 4. where $\delta$ parametrizes the dimerization of the 
chain and $\Delta$ the staggered on-site potential. Inversion symmetry is broken when $\delta\neq 0$ and  
$\Delta\neq 0$ and preserved otherwise. The unit cell along $z$-(of length $c$) has two sites. Since we are interested 
in a model with the minimal number of parameters that captures the physics of the monochalcogenides, we set the distance 
between atoms to be the same and modulate only the hopping. We obtain $\hat{H}=\sum_k (c_{k,A}^{\dagger}~ c_{k,B}^{\dagger} ) H(k) (c_{k,A}~ c_{k,B} )^{T}$ with Bloch  Hamiltonian, 

\begin{align}
H= \sigma_x~ t\cos (k a/2)  - \sigma_y~ \delta \sin (ka/2)  + \sigma_z ~\Delta
\end{align}
and cell periodic functions $u_n$ such that $H u_{c,v} = \pm E u_{c,v}$.  The eigenvalues are given by 
$E=\sqrt{t^2 \cos^2 ka/2 + \delta^2 \sin^2 ka/2 + \Delta^2}$ for the conduction and $-E$ for the valence bands. Berry connections 
will depend explicitly on the gauge used but the results on the shift vector and shift current are gauge independent. We then 
compute the shift current and polarization from the Berry connections and velocity matrix elements. If the electric field is 
along the chain, the $z$-direction, the shift current is  
\begin{align}
J^{z}_{\textrm{shift}}(\omega) = 2 \sigma^{zzz}(0;\omega,-\omega) E^z(\omega)E^z(-\omega).
\end{align}
For the two-band model
\begin{align}
\sigma^{zzz} = e^3 \int_0^{2\pi/a} d k~ \frac{|\langle u_c |v| u_v \rangle|^2 R_{cv} }{\hbar^2\omega^2}  \delta(\frac{2E}{\hbar} -\omega)
\end{align}
To quantify the amount of shift current generated in short-circuit mode we define the ``average conductivity tensor'' as,
\begin{align}
\int d\omega~\sigma^{zzz}(0;\omega,-\omega) = e^3 \int d k \frac{|\langle u_c |v| u_v \rangle|^2  R_{cv}}{4 E^2}
\end{align}
This integral is shown in the main text. It can be expressed analytically in terms of elliptic functions~\cite{Fregoso}. 
Here chose model parameters $(t,[0,\delta_0],\Delta_0)= (1,[0,-.865],.4)$ eV that fit the ab-initio shift current data 
for \textrm{GeS}. In Fig. 3 in the main text we normalized the vertical and horizontal axis with respect to the results 
at the ground state $(t,\delta_0,\Delta_0)= (1,-.865,.4)$. For the other monochalcogenides, we estimate the $\delta_0$ 
parameter by fitting the DFT direct gap (shown in Table 1 in main text) and the polarization curves shown in 
Fig.~\ref{fig:polarization}, i.e., we solve two equations simultaneously,
\begin{align}
E_g  &= \sqrt{\delta_0^2 + \Delta_0^2} \nonumber\\
P_0^{\textrm{1D}} &=  \frac{1}{2\pi}\int dk A_{vv} (k,\delta_0,\Delta_0) -\frac{1}{2\pi}\int dk A_{vv} (k,0^{+},\Delta_0)
\label{eq:pol_1d}
\end{align}
We normalize the $y$ axis in Figs.~\ref{fig:polarization} by $P_0$, and then find the best fit of Eqs.~\ref{eq:pol_1d} 
properly normalized too. In this way, we obtain the $\delta$ and $\Delta$ parameters for the 2D monochalcogines, shown 
in Table.~\ref{table:delta-RM}. As mentioned in the main text, for GeS, GeSe and SnS $\delta_0$ ($=  -0.9$, $-0.5$ and 
$-0.6$ respectively) is relatively close to the optimal $\delta_m$ of $-0.5$, $-0.5$ and $-0.6$ respectively, whereas 
for SnSe $\delta_m$ ($=0.4$) is farthest from $\delta_0=-0.2$.

\begin{table}[t]
\begin{tabular}{lcccccccccccc}
\toprule
      &\hspace*{0.05in}& $P_0^{\textrm{3D}}$ (C/m$^2$) && $E_g$ (eV) &
\hspace*{0.05in}& $\Delta_0/t$  &\hspace*{0.15in}& $\delta_0/t$ &\hspace*{0.15in}& $\delta_m/t$ \\ 
 GeS  && 1.9 &  & 1.9 && 0.45  && -0.83 && -0.52\\ 
 GeSe && 1.3 &  & 1.2 && 0.39  && -0.45 && -0.45\\
 SnS  && 0.8 &  & 1.4 && 0.54  && -0.52 && -0.60\\ 
 SnSe && 0.6 &  & 0.9 && 0.36  && -0.20 && -0.44 \\ 
\bottomrule
\end{tabular}
\caption{RM model parameters for the 2D monochalcogenides. $\delta_0$ and $\Delta_0$ are set to yield the 
DFT band gap $E_g$ and the shapes of the polarization curves in Fig.~\ref{fig:polarization} (read text).}
\label{table:delta-RM}
\end{table}

\LTcapwidth=0.45\textwidth 
\begin{longtable}{lcccclcccclccc}
\caption{Geometry (in {\AA}) of single-layer monochalcogenides used to compute the spontaneous 
polarization along an adiabatic path connecting $\v{R}^i_f(\lambda=-1)$ to $\v{R}^i_f(\lambda=1)$, read manuscript.}
\label{table:coords}
\endfirsthead
\multicolumn{9}{l}{\bf \textrm{GeS}}\\
\multicolumn{9}{l}{Lattice parameters:}\\
$\vec{a} =$ & 15.00 & 0.00 & 0.00 \\
$\vec{b} =$ &0.00   & 3.66 & 0.00 \\   
$\vec{c} =$ &0.00   & 0.00 & 4.47 \\
\multicolumn{9}{l}{Atom coordinates:}\\
&\multicolumn{3}{c}{$R_0$} & 
\hspace{0.1in} &
\multicolumn{3}{c}{$R_f(\lambda=1.0)$}&
\hspace{0.1in} &
\multicolumn{3}{c}{$R_f(\lambda=-1.0)$}\\
\textrm{Ge} & 
 1.41 & 0.91 & 0.00 &&
 1.41 & 0.91 & 0.59 &&
 1.41 & 0.91 & -0.59\\
\textrm{Ge} & 
 3.97 & 2.74 & 2.24&&
 3.97 & 2.74 & 2.83&&
 3.97 & 2.74 & 1.64\\
\textrm{S}  & 
 3.76 & 0.91 & 0.00&&
 3.76 & 0.91 & 0.00&&
 3.76 & 0.91 & 0.00\\
\textrm{S}  & 
 1.62 & 2.74 & 2.24&&
 1.62 & 2.74 & 2.24&&
 1.62 & 2.74 & 2.24\\
\multicolumn{9}{l}{\bf \textrm{GeSe}}\\
$\vec{a} =$ &15.00& 0.00 & 0.00 \\
$\vec{b} =$ &0.00 & 3.98 & 0.00 \\   
$\vec{c} =$ &0.00 & 0.00 & 4.26 \\
\multicolumn{9}{l}{Atom coordinates:}\\
& \multicolumn{3}{c}{$R_0$} & 
\hspace{0.1in} &
\multicolumn{3}{c}{$R_f(\lambda=1.0)$}&
\hspace{0.1in} &
\multicolumn{3}{c}{$R_f(\lambda=-1.0)$}\\
\textrm{Ge} & 
 1.58 & 0.99 & 0.00 &&
 1.58 & 0.99 & 0.44 &&
 1.58 & 0.99 & -0.44\\
\textrm{Ge} & 
 4.00 & 2.98 & 2.13&&
 4.00 & 2.98 & 2.58&&
 4.00 & 2.98 & 1.68\\
\textrm{Se} & 
4.10 & 0.99 &  0.00&&
4.10 & 0.99 &  0.09&&
4.10 & 0.99 & -0.09\\
\textrm{Se} & 
1.49 & 2.98 & 2.13&&
1.49 & 2.98 & 2.22&&
1.49 & 2.98 & 2.04\\
\multicolumn{9}{l}{\bf \textrm{SnS}}\\
\multicolumn{9}{l}{Lattice parameters:}\\
$\vec{a} =$ &15.00& 0.00 & 0.00 \\
$\vec{b} =$ &0.00 & 4.11 & 0.00 \\   
$\vec{c} =$ &0.00 & 0.00 & 4.27 \\
\multicolumn{9}{l}{Atom coordinates:}\\
& \multicolumn{3}{c}{$R_0$} & 
\hspace{0.1in} &
\multicolumn{3}{c}{$R_f(\lambda=1.0)$}&
\hspace{0.1in} &
\multicolumn{3}{c}{$R_f(\lambda=-1.0)$}\\
\textrm{Sn} & 
 1.45 & 1.03 & 0.00 &&
 1.45 & 1.03 & 0.47 &&
 1.45 & 1.03 & -0.47\\
\textrm{Sn} & 
 4.29 & 3.08 & 2.13 &&
 4.29 & 3.08 & 2.61&&
 4.29 & 3.08 & 1.66\\
\textrm{S}  & 
 4.04 & 1.03 & 0.00&&
 4.04 & 1.03 & 0.19&&
 4.04 & 1.03 & -0.19\\
\textrm{S}  & 
 1.71 & 3.08 & 2.13&&
 1.71 & 3.08 & 2.32&&
 1.71 & 3.08 & 1.94\\
\multicolumn{9}{l}{\bf \textrm{SnSe}}\\
\multicolumn{9}{l}{Lattice parameters:}\\
$\vec{a} =$ &15.00 & 0.00 & 0.00 \\
$\vec{b} =$ &0.000 & 4.31 & 0.00 \\   
$\vec{c} =$ &0.000 & 0.00 & 4.38 \\
\multicolumn{9}{l}{Atom coordinates:}\\
& \multicolumn{3}{c}{$R_0$} & 
\hspace{0.1in} &
\multicolumn{3}{c}{$R_f(\lambda=1.0)$}&
\hspace{0.1in} &
\multicolumn{3}{c}{$R_f(\lambda=-1.0)$}\\
\textrm{Sn} & 
 1.58 & 1.08 & 0.00 &&
 1.58 & 1.08 & 0.41 &&
 1.58 & 1.08 & -0.41\\
\textrm{Sn} & 
4.33 & 3.23 & 2.19 &&
4.33 & 3.23 & 2.60 &&
4.33 & 3.23 & 1.78\\
\textrm{Se} & 
 4.31 & 1.08 & 0.00&&
 4.31 & 1.08 & 0.22&&
 4.31 & 1.08 & -0.22\\
\textrm{Se} & 
 1.61 & 3.23 & 2.19&&
 1.61 & 3.23 & 2.41&&
 1.61 & 3.23 & 1.97\\
\end{longtable}

\end{document}